\documentclass[trackchanges]{aastex7}
\usepackage{newtxtext,newtxmath}
\usepackage{babel}
\usepackage{CJKfntef,CJK}
\usepackage[T1]{fontenc}
\usepackage{verbatim}
\usepackage{color}
\usepackage{url}
\usepackage{subcaption}
\usepackage{natbib}
\usepackage{hyperref}
\usepackage{graphicx}	% Including figure files
\usepackage{amsmath}	% Advanced maths commands
\usepackage{siunitx}
\usepackage{threeparttable}
\usepackage{ulem}
\usepackage[dvipsnames]{xcolor}
\usepackage{array}
\usepackage{multirow}

\definecolor{jh}{HTML}{ffc773}

\DeclareRobustCommand{\VAN}[3]{#2}
\let\VANthebibliography\thebibliography
\def\thebibliography{\DeclareRobustCommand{\VAN}[3]{##3}\VANthebibliography}

\newcommand{\ssst}{\scriptscriptstyle}
\newcommand{\E}[1]{\times 10^{#1}}

\newcommand{\s}{\,{\rm s}}      \newcommand{\ps}{\,{\rm s}^{-1}}
    
    \newcommand{\km}{\,{\rm km}}

\newcommand{\erg}{\,{\rm erg}}        \newcommand{\K}{\,{\rm K}}
    
    \newcommand{\G}{\,{\rm G}}

\newcommand{\VLSR}{V_{\ssst\rm LSR}}

\newcommand{\snr}{Kes\,78}
\newcommand{\twCO}{$^{12}$CO}   \newcommand{\thCO}{$^{13}$CO}

\newcommand{\Jotz}{$J$=1--0}

\begin{document}
\begin{CJK*}{UTF8}{bsmi}
\title{Fermi-LAT Detected Gamma-ray Emission Likely Associated with SNR \snr}

%% In v7 the \author command takes an optional argument which provides 
%% additional metadata about the author. Authors can provide the 16 digit 
%% ORCID, the surname (family or last) name, the given (first or fore-) name, 
%% and a name suffix, e.g. "Jr.". The syntax is:
%%
%% \author[orcid=0000-0002-9072-1121,gname=Gregory,sname=Schwarz]{Greg Schwarz}
%%
%% This name metadata in not shown, it is only for parsing by the peer review
%% system so authors can be more easily identified. This name information will
%% also be sent to the publisher so they can include it in the CROSSREF 
%% metadata. Including an orcid will hyperlink the author name to the 
%% author's ORCID page. Note that  during compilation, LaTeX will do some 
%% limited checking of the format of the ID to make sure it is valid. If 
%% the "orcid-ID.png" image file is  present or in the LaTeX pathway, the 
%% ORCID icon will appear next to the authors name.
%%
%% Even though emails are now required for each author, the \email does not
%% produce output in the compiled manuscript unless the optional "show" command
%% is used. For example,
%%
%% \email[show]{greg.schwarz@aas.org}
%%
%% All "shown" emails are show in the bottom left of the first page. Due to
%% space constraints, only a few emails should be shown. 
%%
%% To identify a corresponding author, use the \correspondingauthor command.
%% The command appends "Corresponding Author: " to the argument it appears at
%% the bottom left of the first page like the output from \email. 

\author[orcid=0009-0006-5656-6902, sname='Shen', gname='Yun-Zhi']
{Yun-Zhi Shen(沈蘊之)}
\affiliation{School of Astronomy \& Space Science, Nanjing University, 163 Xianlin Avenue, Nanjing 210023, China}
\email[]{yzshen@smail.nju.edu.cn}

\author[0000-0002-4753-2798]
{Yang Chen(陳陽)}
\affiliation{School of Astronomy \& Space Science, Nanjing University, 163 Xianlin Avenue, Nanjing 210023, China}
\affiliation{Key Laboratory of Modern Astronomy and Astrophysics, Nanjing University, Ministry of Education, Nanjing 210023, China}
\email[]{ygchen@nju.edu.cn}

\author[0000-0002-9392-547X]{Xiao Zhang(張瀟)}
\affiliation{School of Physics \& Technology, Nanjing Normal University, No.1 Wenyuan Road, Nanjing 210023, China}
\email[]{xiaozhang@njnu.edu.cn}

\author[0000-0003-4916-4447]{Chen Huang({黃晨})}
\affiliation{School of Astronomy \& Space Science, Nanjing University, 163 Xianlin Avenue, Nanjing 210023, China}
\email[]{chenhuang@smail.nju.edu.cn}
\correspondingauthor{Yang Chen \& Xiao Zhang}
\email{ygchen@nju.edu.cn; xiaozhang@njnu.edu.cn}
%\collaboration{all}{The Terra Mater collaboration}

\begin{abstract}
We have analyzed the GeV gamma-ray emission in the region of the supernova remnant (SNR) \snr\ using $\sim$16.7 years of {\sl Fermi}-LAT observations and found that the catalog sources 4FGL J1852.4+0037e and 4FGL J1851.8$-$0007c are better represented as two extended sources modeled as {\sl 2Ext}.
One of them, designated as E2, is located at R.A.$=282.86^\circ$, Dec.$=-0.11^\circ$ with the 68\% containment radius $R_{68} = 0.31^\circ$, and is detected with a significance of 15.2$\sigma$ in the 0.2--500 GeV energy range. 
The gamma-ray emission of source E2 is well described by a log-parabola (LogP) spectral model with spectral index $\Gamma$ = 1.2 and curvature $\beta$ = 0.3. 
The fitting with electron-proton number ratio $K_{\rm ep}=0.01$ indicates that the GeV emission of source E2 is dominated by hadronic emission.
Given the dense molecular environment surrounding the middle-aged SNR \snr, the hadronic scenario provides a natural explanation for the observed GeV emission.
The extended source E2 can also be replaced with two point sources. 
One of them, designated as PTS1, is coincident with the newly discovered PSR J1852$-$0002g within the 68\% positional uncertainty circle, indicating a possible gamma-ray contribution from this PSR. 
The gamma-ray spectrum of source PTS1 can be well described by a LogP spectral shape. 
The synchro-curvature radiation model provides a satisfactory spectral fit for source PTS1, suggesting that some of the GeV emission from the \snr\ region might possibly originate from the magnetosphere of PSR J1852$-$0002g. 
%However, no gamma-ray pulsation has been found from the pulsar in our work, possibly due to limited observational sensitivity or accurate ephemeris.

%We present the detection of an extended GeV gamma-ray source likely associated with the SNR \snr, based on an analysis of approximately 16.7 years of {\sl Fermi}-LAT data. 
%The source, designated E2, exhibits spatial coincidence with the SNR and is detected with a significance of 15.2$\sigma$ in the 0.2--500\,GeV energy range {with log-parabola (LogP) spectral type. 
%We use the number ratio between accelerated electrons and protons $K_{\rm ep} = 0.01$ to fit the SED of source E2.
%The GeV emission of source E2 can be naturally explained by hadronic process, given that SNR \snr\ is evolving within a dense molecular environment.
%This extended source E2 can also be replaced with two point sources, 
%Although the extended source in our spatial model is the best-fitted case, the emission in this region can be modeled as two point sources with moderately lower TS value.
%one of them, designated as PTS1, appears spatially coincident with PSR J1852$-$0002g within the 68\% positional uncertainty circle of source PTS1 with LogP type.
%, with a photon index of approximately 2.0. 
%In addition, we separately fit the SED of source PTS1 with the synchro-curvature radiation model, which implies a possible contribution from the PSR J1852$-$0002g in \snr\ region.

\end{abstract}



\section{Introduction} 
%\red{The primary sources of GeV emission within the Galaxy include contributions from supernova remnants (SNRs) and pulsars (PSRs). }
%Since pulsars (PSRs) have been observed to emit pulsed radiation above 20 TeV \citep[e.g., Vela;][]{HESS_vela_20tev}, they are increasingly recognized, alongside supernova remnants, as important Galactic sources of particle acceleration.
Supernova remnants (SNRs) are widely regarded as prominent Galactic particle accelerators, capable of accelerating particles to energies of several hundred TeV.
Observations from the {\sl Fermi}-LAT have shown that the GeV gamma-ray emission from middle-aged SNRs is typically attributed to hadronic processes, where gamma-rays are produced via the decay of neutral pions created in proton-proton interactions \citep[e.g., W51C, W28, W44;][]{W51C_midage, W28_highB, w28_cui_2018, W44_midage, W44_Peron_2020}.
Multi-wavelength studies are consistent with or further support this interpretation by revealing spatial correlations between the gamma-ray emission and the adjacent molecular clouds (MCs) associated with the SNRs \citep[e.g.,][]{kes41, G51.2, Kes67_GeV, G298.6-0.0}.   
%\red{Supporting evidence for the interaction between the SNR and MCs includes the detection of 1720 MHz OH masers and the spatial correlation between the SNR and nearby MCs \citep{Jiang}, which can help constrain the hadronic origin of the observed GeV gamma-ray emission.}
%This hadronic interaction is characterized by a distinctive spectral signature known as the "pion-decay bump" \citep{pp_bump}. 
%Additional evidence for SNR-MC interactions includes the detection of 1720 MHz OH masers, \red{the morphological agreement between SNR and MCs \citep{Jiang} serving as key indicators of hadronic} processes and helping to constrain the origin of the observed GeV gamma-ray emission.
In some cases, there are both SNR and pulsar (PSR) in a GeV gamma-ray emission region.
The GeV gamma-ray emission from PSRs originates from high-energy processes occurring within their magnetospheres, primarily through synchrotron and curvature radiation mechanisms \citep{PSR_book}. 
%In {\sl Fermi}-LAT observations, pulsar emissions typically manifest as point sources. 
%Theoretical advances have led to a unified model that combines these two mechanisms into synchro-curvature (SC) radiation \citep{syn-cur-1996, syn-chr_emission_formulae}. 
%This model has been successfully applied to explain the gamma-ray emission from well-known pulsars such as Geminga, Crab, and Vela \citep{syn-chr-fit}. 
%Moreover, SC radiation has also been used to interpret the GeV–TeV emission from sources without clear counterparts, such as LHAASO J0314+5258 \citep{syn-chr_LHA0341}.
This study focuses on the gamma-ray emission from the SNR–MC interaction region of SNR Kesteven\,78 (G32.8$-$00.1; hereafter \snr), aiming to assess the potential contribution from hadronic processes.
Meantime, the recent discovery of PSR J1852$-$0002g, located to the east of the SNR shock boundary, introduces the possibility of an additional contribution from pulsar-related emission.
%for the existence of PSR J1852$-$0002g, \red{which is identified by FAST \citep{FAST_find_PSR} with the spin period of 0.24510 s and the distance of 5.6 kpc, to the east of SNR shock boundary.}
%Moreover, \citet{FAST_find_PSR} has identified a radio pulsar located east of \snr\ with the spin period of 0.24510 s and the distance of 5.6 kpc.

SNR \snr\ exhibits an ellipse-like morphology in both radio and X-ray bands, with its structure appearing elongated along the north–south orientation and compressed in the east–west (J2000).
A 1720 MHz OH maser has been detected along the eastern boundary of \snr, at coordinates (18$^{\rm h}$51$^{\rm m}$48$^{\rm s}$.04, $-$00$^\circ$10$'$35$''$; J2000) and a local standard of rest (LSR) velocity $V_{\rm LSR}$ of +86.1 $\km\ps$ \citep{OH_maser}, providing robust evidence for shock interaction between the SNR and surrounding MCs.
%Previous studies have shown that \snr\ is evolving within a dense environment and have identified additional indicators of SNR-MC interaction, including broadene d \twCO\ line profiles and elevated \twCO(\Jtto)/(\Jotz) ratios along the remnant’s periphery \citep{Zhou_kes78_2011}.
The SNR has been suggested to be associated with molecular gas at a systemic $V_{\rm LSR}$ of approximately +81~$\km\ps$ \citep{Zhou_kes78_2011}.
The kinematic distance to the SNR is estimated as 4.8 kpc based on the $V_{\rm LSR} \sim$ +81 $\km\ps$ for both CO \citep{Zhou_kes78_2011} and HI \citep{distance_HI} observations, and $\sim$ 5.4 kpc for H$_2$ emission line method with the $V_{\rm LSR}$ of +90 $\km\ps$ \citep{H2_distance}.
We will parameterize the distance of SNR \snr\ as $d=d_5$ kpc.
\citet{Zhou_kes78_2011} has reported the underionized hot ($\sim$ 1.5 keV), low-density ($\sim$ 0.1 cm$^{-3}$) inter-cloud plasma and inferred the age of the remnant as about 6 kyr.
X-ray observations with {\sl Suzaku} suggest that \snr\ is a middle-aged remnant with an estimated age of approximately $2.2 \times 10^4$ yr, indicating an average shock velocity of 500 -- 700 $\km\ps$ \citep{Kes78_suzaku}.
{\sl XMM-Newton} observations reveal that the eastern region of \snr\ contains cooler and denser plasma than that in the western region, which may signal interaction between the SNR shock and adjacent MCs \citep{Kes78_XMM}.
To the east of SNR \snr, a TeV source HESS J1852$-$000 was detected \citep{HESS_catalog}.
PSR J1852$-$0002g, located to the east of the SNR shock boundary, is identified by FAST \citep{FAST_find_PSR} with a spin period of 0.24510 s at a distance of 5.6 kpc.
\citet{He_Kes78_2022} investigates the distribution of GeV gamma-ray emission in the region surrounding SNR Kes79  using 11.5 years of {\sl Fermi}-LAT observations, identifying two distinct sources, Src-N and Src-S. Notably, Src-S is located to the southeast of SNR Kes\,79 with an angular radius of 0.58$^\circ$, and encompasses SNR \snr.
However, as early as the release of the 4FGL-DR2 catalog, based on 10 years of data,  a gamma-ray source was already reported to the east of SNR \snr's shock boundary \citep{4FGL, 4FGL-DR2}. These detections motivate a dedicated investigation of the GeV gamma-ray emission potentially associated with SNR \snr.
%\red{PSR J1852$-$0002g, located to the east of the SNR shock boundary, is identified by FAST \citep{FAST_find_PSR} with a spin period of 0.24510 s and a distance of 5.6 kpc.}
%While \citet{Kes78_suzaku} report the presence of an additional hard X-ray component in the {\sl Suzaku} spectrum, \citet{Kes78_XMM} find no evidence for synchrotron emission in the {\sl XMM-Newton} data.
%No further timing parameters are currently available.

In this work, we resolve the GeV gamma-ray emission associated with SNR \snr\ and also possibly PSR J1852$-$0002g and explore the origin of the emission. By analyzing the 16.7 yr {\sl Fermi}-LAT data, we resolve an extended source overlapping \snr\ and fit the spectral energy distribution (SED) to examine the emission mechanisms. 
We also employ the synchro-curvature model to study the possible contribution from PSR J1852$-$0002g.
The observational data are described in Section~\ref{sec:observation} and the analysis results are presented in Section~\ref{sec: results}. The results are discussed in Section~\ref{sec:discussion}.

\section{Observations and data}\label{sec:observation}

\subsection{{\sl Fermi}-LAT Observation Data}
For the gamma-ray emission, we use 16.7 yr observation data of Large Area Telescope\,(LAT) onboard the {\sl Fermi Gamma-ray Space Telescope}. The time frame of our research is from 2008-08-04 15:43:36 (UTC) to 2025-04-02 00:50:50 (UTC), and the circular region of interest (ROI) is 15$^{\circ}$ in radius, centered at the coordinates R.A.=282.97$^{\circ}$, Dec=$-$0.12$^{\circ}$ (J2000).
We use the software {\small Fermipy}\footnote{\url{https://fermipy.readthedocs.io/en/stable/}} (Vertion 1.3.1 released on 2024 August 20), which is based on the {\small Fermitools} \footnote{\url{http://fermi.gsfc.nasa.gov/ssc/data/analysis/software/}}(Version 2.2.0 realesed on 2022 June 21), to analyze the data.
We select `SOURCE' class (evclass=128, evtype=3) with the instrument response function (IRF) `P8R3\_SOURCE\_V3\_v1' and constrain the energy range to 0.2 -- 500\,GeV. To eliminate the Earth's limb, we limit the maximum of zenith to 90$^{\circ}$. 
We also apply the recommended filter string `(DATA\_QUAL\textgreater0)\&\&(LAT\_CONFIG==1)' to choose the good time intervals. 
We construct the background model by integrating gamma-ray sources from the {\sl Fermi}-LAT Fourth Source Catalog Data Release 4 \citep[4FGL-DR4, incorporating updates from 4FGL-DR3;][]{4FGL-DR4, 4FGL-DR3} with the ROI to study the gamma-ray sources, with the Galactic diffuse emission modeled using \textit{gll\_iem\_v07.fits} and isotropic emission (\textit{iso\_P8R3\_SOURCE\_V3\_v1.txt}).
To enable a refined spatial analysis, we exclude the PSF0 data and reselect the `CLEAN' class (evclass=256, evtype=56)\footnote{\url{https://fermi.gsfc.nasa.gov/ssc/data/analysis/documentation/Cicerone/Cicerone_Data/LAT_DP.html}\label{data mode}} using the IRF `P8R3\_CLEAN\_V3::PSF' \footnote{\url{https://fermi.gsfc.nasa.gov/ssc/data/access/lat/BackgroundModels.html}}.
This IRF is associated with the isotropic background model corresponding to each PSF type (e.g., \textit{iso\_P8R3\_CLEAN\_V3\_PSF\_v1.txt} for PSF1).

\subsection{Molecular Line Data}
We use the archival data of the \twCO\ \,(\Jotz) line at \SI{115.271}{GHz} and the \thCO\ \,(\Jotz) line at \SI{110.201}{GHz} of the FOREST Unbiased Galactic plane Imaging survey with the Nobeyama 45 m telescope (FUGIN; \citet{FUGIN}) observation. 
The angular resolution was 20$^{\prime\prime}$ for \twCO\ and 21$^{\prime\prime}$ for \thCO, and the average rms noise was $\sim$\SI{1.5}{\kelvin} for \twCO\ and $\sim$\SI{0.7}{\kelvin} for \thCO\ at a velocity resolution of 0.65\,km\,s$^{-1}$. 

\subsection{Other Data}
We use the SARAO MeerKAT 1.3 GHz Galactic Plane Survey (SMGPS) continuum image with an angular resolution of 8$''$ \citep{MeerKAT} to delineate the radio brightness distribution of the SNR.

\section{{\sl Fermi}-LAT GeV gamma-ray emission analysis}\label{sec: results}

\subsection{Spatial Analysis}

\begin{figure}

    \includegraphics[width=1.0\textwidth]{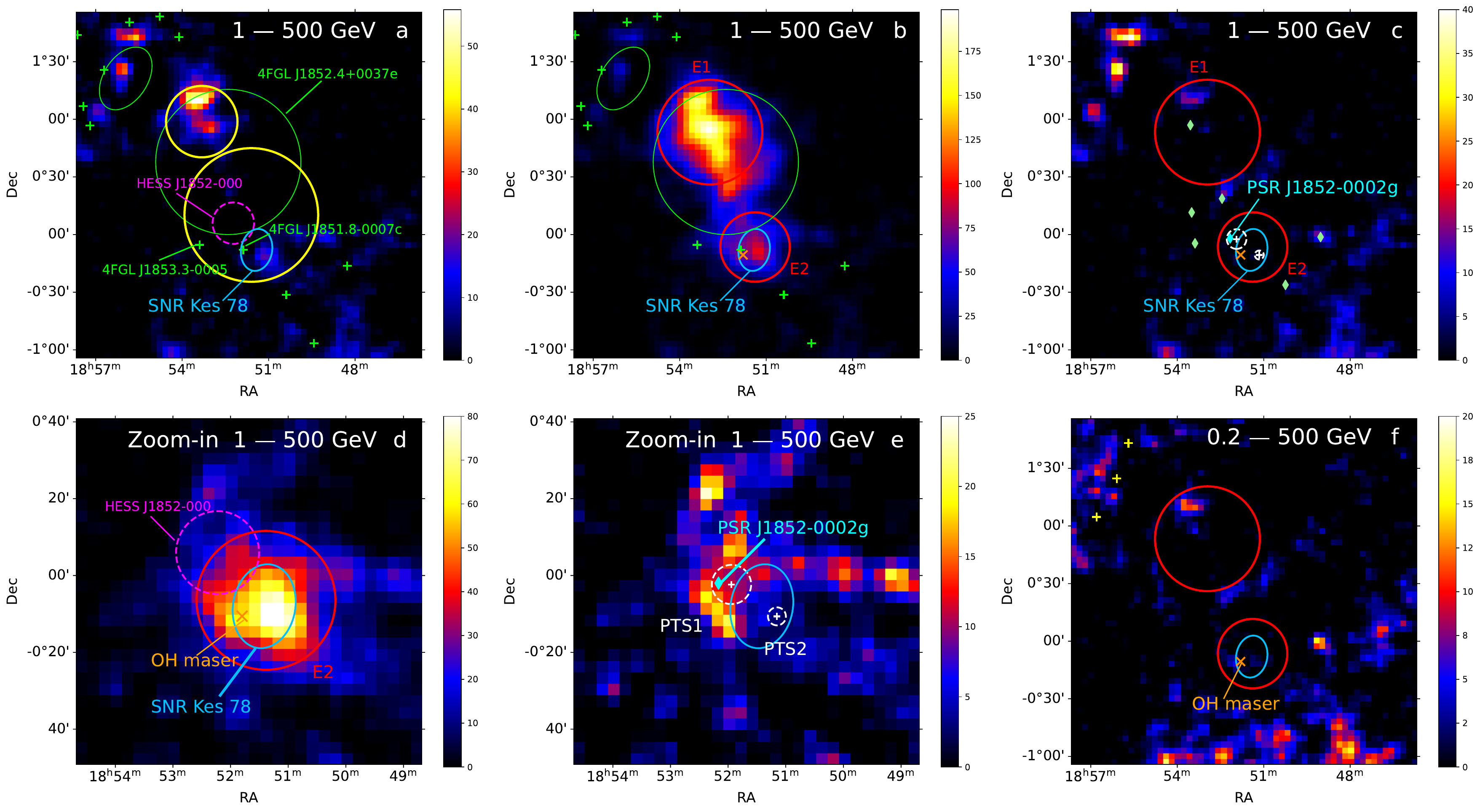}
    \caption{TS maps of 3$^{\circ}$$\times$3$^{\circ}$ field covering SNR Kes\,78 of {\sl Fermi}-LAT in 1 - 500 GeV and 0.2 - 500 GeV. The image scale of the TS maps is 0.05$^{\circ}$ per pixel. 
    {\sl a)} TS map of 4FGL-DR4 catalog model. 
    {\sl b)} TS map of catalog model without 4FGL J1852.4+0037e and 4FGL J1851.8$-$0007c in the spatial model. 
    {\sl c)} Residual TS map of spatial model {\sl 2Ext} in the energy range of 1 -- 500 GeV, with the maximum residual value within source E1 equal to 15.7. 
    {\sl d, e)} Zoomed in TS distribution of sources `E2' and `PTS1' of 1.5 $^\circ \times$ 1.5$^\circ$ field, respectively.
    {\sl f)} Residual TS map of spatial model {\sl 2Ext} in the energy range of 0.2 -- 500 GeV, with the maximum residual value within source E1 equal to 11.8.
    The yellow circles in panel {\sl a} are the extension of sources Src-N and Src-S in \citet{He_Kes78_2022}.
    The green circle and crosses in panels {\sl a} and {\sl b} are the 4FGL-DR4 catalog sources and the two red circles indicate the 68\% containment radii of sources in the spatial model {\sl 2Ext}. 
    The dashed magenta circles in panels {\sl a} and {\sl d} indicate the TeV source HESS J1852$-$000.
    The blue ellipse represents the position of SNR \snr, and the orange crosses in panels {\sl c} and {\sl f} mark the position and spatial extent of the 1720 MHz OH maser. 
    The light-green diamonds in panel {\sl c} mark the PSRs with the spin-down energy above $1\E{34} \erg\ps$ and the cyan diamond marks the PSR J1852$-$0002g. 
    The dashed white circles and crosses in panels {\sl c} and {\sl e} are the point sources in {\sl 1Ext 2PS} model listed in Table ~\ref{tab: spatial models}.
    The yellow crosses in panel {\sl f} mark the sources added in the background spatial model, which are listed in Table~\ref{tab: ps sources}.}
    \label{fig: TSmaps}
\end{figure}

In the vicinity of SNR \snr, two sources, 4FGL J1851.8$-$0007c and 4FGL J1852.4+0037e, are listed in the 4FGL-DR4 catalog.
%To re-examine the morphology of the GeV gamma-ray emission in this region, we exclude these two catalog sources from the background model.
To reduce statistical uncertainties, we select photon events with energies above 1\,GeV and exclude the PSF0\footref{data mode} data. 
In the spatial model fitting, we allow the spectral parameters of the sources within a 5$^\circ$ radius from the center of ROI to vary freely, along with the parameters of the Galactic diffuse and isotropic background components.
We then generate residual test statistic (TS) maps 
%over a 3$^\circ \times$ 3$^\circ$ region 
centered on the ROI center for both the catalog model and the model in which the two aforementioned sources are removed.
The TS is defined as TS$ = 2\log{\left({\cal L}_1 / {\cal L}_0\right)}$, where ${\cal L}_0$ and ${\cal L}_1$ represent the maximum likelihood values under the null hypothesis and the alternative hypothesis (with a test source at each pixel), respectively.
As shown in Figure~\ref{fig: TSmaps}a, for the catalog model, residual gamma-ray emission remains in the northeastern region of the catalog source 4FGL J1852.4+0037e. In contrast, Figure~\ref{fig: TSmaps}b, the residual map without subtracting the emission of catalog sources 4FGL J1852.4+0037e and 4FGL J1851.8$-$0007c reveals a distinct gamma-ray emission peak in the northeast in the map and a structure extending southwestward from the peak, which cannot be well modelled by the catalog sources.

We therefore reanalyze the spatial distribution of GeV gamma-ray emission in this region. Using the \texttt{extension} method provided by {\small Fermipy}, we refit the catalog source 4FGL J1852.4+0037e to determine its best-fit position and spatial extension. 
Additionally, we apply the \texttt{localize} method to refit the spatial parameters of the catalog source 4FGL J1851.8$-$0007c and the \texttt{extension} method to test its spatial extension.
The 68\% containment radius, $R_{68}$, is adopted to describe the extension of extended sources.
%Table~\ref{tab: spatial models} lists the parameters of the refitted spatial models.
The extension significance is quantified by the test statistic TS$_{\rm ext} = 2\log{(\mathcal{L}_{\rm ext} / \mathcal{L}_{\rm PS})}$, where \(\mathcal{L}_{\rm ext}\) and \(\mathcal{L}_{\rm PS}\) are the maximum likelihoods of the extended and point-source hypotheses, respectively. A source is considered extended if TS$_{\rm ext} > 16$.
We also compare two spatial templates for modeling extended sources: a uniform disk and a radial Gaussian profile. We find the difference of their likelihood values is less than 2, which indicates no significant differences in the fitting results between the two templates, and thus adopt the Gaussian template for subsequent analyses.
%As a result, we find that 4FGL J1851.8$-$0007c is significantly extended with TS$_{\rm ext} = 34.0$. The best-fit spatial parameters are: R.A. = 282.86$^\circ$ $\pm$ 0.03$^\circ$, Decl. = $-$0.11$^\circ$ $\pm$ 0.03$^\circ$ (J2000), with the 68\% containment radius ($R_{68}$) of 0.31$^\circ$ $\pm$ 0.04$^\circ$.
%The refitted source 4FGL J1852.4+0037e is labeled `E1', and the refitted source 4FGL J1851.8$-$0007c is labeled `PTS1' in the {\sl catalog refit} model and `E2' in the {\sl 2Ext} model. 
To further confirm the extended nature of a source, we follow the approach of \citet{TS_ext_compare_2PS} by comparing an extended source model with a two point-source model. The comparison is quantified using the test statistic TS$_{\rm 2PTS} = 2\log{(\mathcal{L}_{\rm 2PTS}/\mathcal{L}_{\rm PS})}$, where $\mathcal{L}_{\rm 2PTS}$ and $\mathcal{L}_{\rm PS}$ represent the likelihoods of the two point-source model in {\sl 1Ext 2PS} and single point-source model in {\sl 1Ext 1PS}, respectively.

In the following we use four spatial models to fit the gamma-ray emission of catalog sources 4FGL J1851.8$-$0007c and 4FGL J1852.4+0037e. In the model names, {\sl Ext} is the notation of extended source, which is chosen as the Gaussian template, and {\sl PS} stands for point source.

\begin{enumerate}
    \item {\sl catalog (1Ext 1PS)} includes the two catalog sources 4FGL J1851.8$-$0007c and 4FGL J1852.4+0037e, as the baseline model.
    \item {\sl catalog refit (1Ext 1PS)} is the refitted catalog model. The refitted source 4FGL J1852.4+0037e is denoted as `E1' and the refitted source 4FGL J1851.8$-$0007c as `PTS1'.
    \item {\sl 2Ext} is the model in that the source `PTS0' is replaced with the extended source `E2', the parameters of which are obtained by applying \texttt{extension} method to 4FGL J1851.8$-$0007c.
    \item {\sl 1Ext 2PS} model includes the extended source `E1' defined in model {\sl 2Ext} and two point sources for a comparison with the extended source `E2' defined in the 2{\sl Ext} model.
\end{enumerate}
For comparison, we also incorporate the spatial model proposed by \citet{He_Kes78_2022}, hereafter referred to as {\sl He2022} model.

%For comparison, we incorporate the spatial model proposed by \citet{He_Kes78_2022}, hereafter referred to as {\sl He2022} model. 
%We evaluate three spatial models in total: the original catalog model, our {\sl 2Ext} model, and the {\sl He2022} model. 
%The spatial parameters for all of the models are summarized in Table~\ref{tab: spatial models}.
%Each model is implemented into the background model with power-law (PL) spectral shape, and the fitting is performed using {\small Fermipy}.
%In the fitting process, we free the spectral parameters of sources within 5$^\circ$ of the ROI center, including those of the target sources as well as the Galactic diffuse and isotropic background components.

\begin{table}
    \centering
    \caption{Best-fitted spatial parameters for all three models.}
    \label{tab: spatial models}
    \begin{tabular}{ccccccc}
         \hline
         \hline
         Model name  & Source Name &  R.A.(J2000)$^a$ ($^\circ$)& Dec.(J2000)$^a$  ($^\circ$) & Extension ($R_{68}$)$^b$ ($^\circ$) &$\bigtriangleup k$ & $\bigtriangleup$AIC\\
         \hline
         {\sl catalog}    & 4FGL J1852.4+0037e 
                                   & 283.1 & 0.63 & 0.63 & \multirow{2}{*}{0} &
                                   \multirow{2}{*}{0}\\
         {\sl (1Ext 1PS)} & 4FGL J1851.8$-$0007c & 282.97 & $-$0.12 & --- \\
         \hline
          {\sl catalog refit}      & E1 & 283.24  $\pm$ 0.02 & 0.89 $\pm$ 0.02
                                   & 0.45$^{+ 0.02}_{- 0.03}$& 
                                   \multirow{2}{*}{0} &
                                   \multirow{2}{*}{$-$136.7} \\
          {\sl (1Ext 1PS)}         & PTS0 & 282.83 $\pm$ 0.02 & $-$0.13 $\pm$ 0.03
                                    & ---  \\
        \hline
         \multirow{2}{*}{{\sl 2Ext}}  & E1 & 283.24 $\pm$ 0.02 & 0.89 $\pm$ 0.02 
                                   & 0.45$^{+ 0.02}_{- 0.03}$& 
                                   \multirow{2}{*}{1} &
                                   \multirow{2}{*}{$-$171.2} \\
                                   & E2 & 282.86 $\pm$ 0.03 & $-$0.11 $\pm$ 0.03
                                   &0.31 $\pm$ 0.04\\
        \hline
        \multirow{3}{*}{{\sl 1Ext 2PS}} & E1  &283.24 $\pm$ 0.02 
                                    & 0.89 $\pm$ 0.02 
                                   & 0.45$^{+ 0.02}_{- 0.03}$& 
                                   \multirow{3}{*}{4} &
                                   \multirow{3}{*}{$-$146.2} \\
                                    & PTS1 & 282.99 $\pm$ 0.09 & $-$0.04 $\pm$ 0.05
                                    & ---  \\
                                    & PTS2 & 282.79 $\pm$ 0.03& $-$0.18 $\pm$ 0.03
                                    & --- \\
        \hline
        {\sl He2022 } & Src-N & 283.33       & 0.98
                                    & 0.31   & \multirow{2}{*}{1} 
                                    & \multirow{2}{*}{$-$129.6} \\
        {\sl \citep{He_Kes78_2022}} & Src-S & 282.90       & 0.17
                                    & 0.58 \\
        \hline
    \end{tabular}
    \begin{tablenotes}
        \small
        \item $^a$ Fitted position coordinates with 1$\sigma$ uncertainty if available.
        \item $^b$ 68\% containment radius with 1$\sigma$ uncertainty if available.
    \end{tablenotes}
\end{table}

\begin{table}
    \centering
    \caption{The extra point sources added in the background model.}
    \label{tab: ps sources}
    \begin{tabular}{ccc}
         \hline
         \hline
         Source  &  R.A.(J2000) ($^\circ$)& Dec.(J2000) ($^\circ$)\\
         \hline
         BPS1     &  284.03              & 1.41\\
         BPS2     &  283.92              & 1.72\\
         BPS3     &  284.20              & 1.08\\
         \hline
    \end{tabular}
\end{table}

The spatial parameters for all of the models are summarized in Table~\ref{tab: spatial models}. Each model is implemented into the source list with power-law (PL) spectral shape, and the fitting is performed using {\small Fermipy}.
To determine the best-fitting spatial model, we adopt the Akaike Information Criterion \citep[AIC;][]{AIC1974, TS_ext_compare_2PS}, defined as AIC = 2$k$ $-$
2$\log{\mathcal{L}}$, where $k$ is the number of free parameters and $\mathcal {L}$ is the maximum likelihood value.
The calculated differences in $\Delta {\rm AIC}$ for the various models are listed in Table~\ref{tab: spatial models}. A more negative AIC value indicates a better-fitting model, and a difference of $|\Delta  {\rm AIC}| > 10$ is considered as strong evidence in favor of the model with the lowest AIC value.
Therefore, the {\sl 2Ext} model, which yields the lowest AIC value, is considered the most favorable and is adopted for subsequent analysis.
For source E2, as a result of \texttt{extension} method, we find that it is significantly extended with TS$_{\rm ext} = 34.0$.
%We replace E2 with two point sources \red{({\sl 1Ext 2PS} model)} and obtain TS$_{\rm 2PTS} = 20.3$. Since TS$_{\rm ext} >$ TS$_{\rm 2PTS}$, E2 is preferred and is treated as an extended source in the subsequent analysis.
Based on these criteria, we adopt the {\sl 2Ext} spatial model for subsequent analyses. 
The best-fit spatial parameters of source E2 are: R.A. = 282.86$^\circ$ $\pm$ 0.03$^\circ$, Dec. = $-$0.11$^\circ$ $\pm$ 0.03$^\circ$ (J2000), with $R_{68}$ of 0.31$^\circ$ $\pm$ 0.04$^\circ$.
As shown in Figure~\ref{fig: TSmaps}c, the residual TS map corresponding to the {\sl 2Ext} model reveals minimal residual gamma-ray emission in the region, indicating a satisfactory fit.
The TS distribution of source E2 is presented in Figure~\ref{fig: TSmaps}d, which shows that the extent of source E2 overlaps SNR Kes 78.
Thus, we focus our subsequent analysis on this source.
%Figure~\ref{fig: TSmaps}d shows the TS distribution of source E2, while Figure~\ref{fig: TSmaps}e shows the emission of source PTS1, which appears to correspond to PSR J1852$-$0002g in position with the 68\% uncertainty.
There are also residual emissions to the northeast of source E1 as seen in Figure~\ref{fig: TSmaps}c. 
Three additional sources with PL spectral models are needed for fitting these remaining residuals in the background.
The spatial parameters of the three sources are listed in Table~\ref{tab: ps sources}. 
As shown in Figure~\ref{fig: TSmaps}f, the inclusion of these sources effectively removes significant residual gamma-ray emission from the background.

Additionally, the extended source E2 spatially overlapping SNR Kes 78 also can be fitted by using two point sources (namely, {\sl 1Ext 2PS} model) with moderately lower significance (TS$_{\rm 2PTS} = 17.5$). 
But we note that the fitted position of source PTS1 is coincident with that of PSR J1852$-$0002g within the 68\% uncertainty range, as can be seen in Figure 1e.
Considering this, we use the \texttt{tempo2} software to search for possible pulsed signals from this PSR, however, no significant detection was obtained.
%, likely due to insufficient or incomplete ephemeris parameters.}
In the following, we will also analyze the property of source PTS1 of the {\sl 1Ext 2PS} model.

%In the context of the {\sl 2Ext} model, we replace E2 with two point sources and obtain TS$_{\rm 2pts} = 20.3$. Since TS$_{\rm ext} >$ TS$_{\rm 2pts}$, E2 is supported to be an extended source in the subsequent analysis. Figure~\ref{fig: TSmaps}c shows that {\sl 2Ext} model can fit the emission well in the region near the catalog sources 4FGL J1852.4+0037e and 4FGL J1851.8$-$0007c.

\subsection{Spectral Analysis}\label{sec: spec}

\begin{table}
    \centering
    \caption{Formulae for gamma-ray spectra}
    \label{tab:formulae}
    \begin{tabular}{ccccccccc}
    \hline
    \hline
    Name &  Formula & Free Parameters & $\Delta k$ & \multicolumn{2}{c}{$\Delta$AIC} & $\Gamma$ & $E_{\rm cut/break}$(GeV) & $\beta$\\
         &          &                 &                 & E2   &  PTS1 &    &     &   \\
    \hline
     PL  & $dN/dE = N_0\left(E/E_0\right)^{-\Gamma}$ 
         & $N_0, \Gamma$ & 0 & 0 & 0 & --- & --- & ---\\
     ECPL& $dN/dE = N_0\left(E/E_0\right)^{-\Gamma}\exp{\left(-E/E_{\rm cut}\right)}$ 
         & $N_0, \Gamma, E_{\rm cut}$ & 1 & $-$17.2 & $-$33.0 & 1.6 & 1.2 & ---\\
     LogP& $dN/dE = N_0\left(E/E_0\right)^{-\Gamma-\beta\log{\left(E/E_0\right)}}$ 
         & $N_0, \Gamma, \beta$ & 1 & $-$25.6 & $-$39.0
         & 1.2 & --- & 0.3\\
     BPL & $dN/dE = N_0\begin{cases}
                    \left(E/E_{\rm b}\right)^{-\Gamma_1}& \text{$E\leq E_{\rm b}$}\\
                    \left(E/E_{\rm b}\right)^{-\Gamma_2}& \text{$E\geq E_{\rm b}$}
                    \end{cases}$
         & $N_0, E_{\rm b}, \Gamma_1, \Gamma_2$ & 2 & $-$20.6 & $-$28.2 & 1.3 (2.5) & 1.0 & --- \\
     \hline
    \end{tabular}
\end{table}

To investigate the spectral properties of source E2 across the full energy range of 0.2 -- 500 GeV using all event types, we test three alternative spectral models including log-parabola (LogP), exponential cutoff power-law (ECPL), and broken power-law (BPL), in addition to the simple power-law (PL) model. The functional forms of these spectral models are summarized in Table~\ref{tab:formulae}.
We also use the AIC method to determine the most appropriate spectral model.
While keeping the spectral type of source E1 fixed as PL, we vary the spectral type of source E2. 
Comparison of the four spectral models as listed in Table~\ref{tab:formulae} shows that the LogP model has the lowest AIC value, and thus it is adopted for refitted 4FGL J1851.8$-$0007c or source E2.
%Following the catalog spectral type of 4FGL J1851.8$-$0007c, we adopt the LogP model with the lowest AIC value.
With the best-fitting LogP spectral model, the parameters of which are $\Gamma=1.2$ and $\beta$ = 0.3, the obtained flux in 0.2 -- 500 GeV is 2.5 $\times\,10^{-11}$ erg cm$^{-2}$ s$^{-1}$, and the luminosity is 7.6 $\times\,10^{34}d_{5}^2$ erg s$^{-1}$.

\begin{figure}
    \centering\includegraphics[width=1.\textwidth]{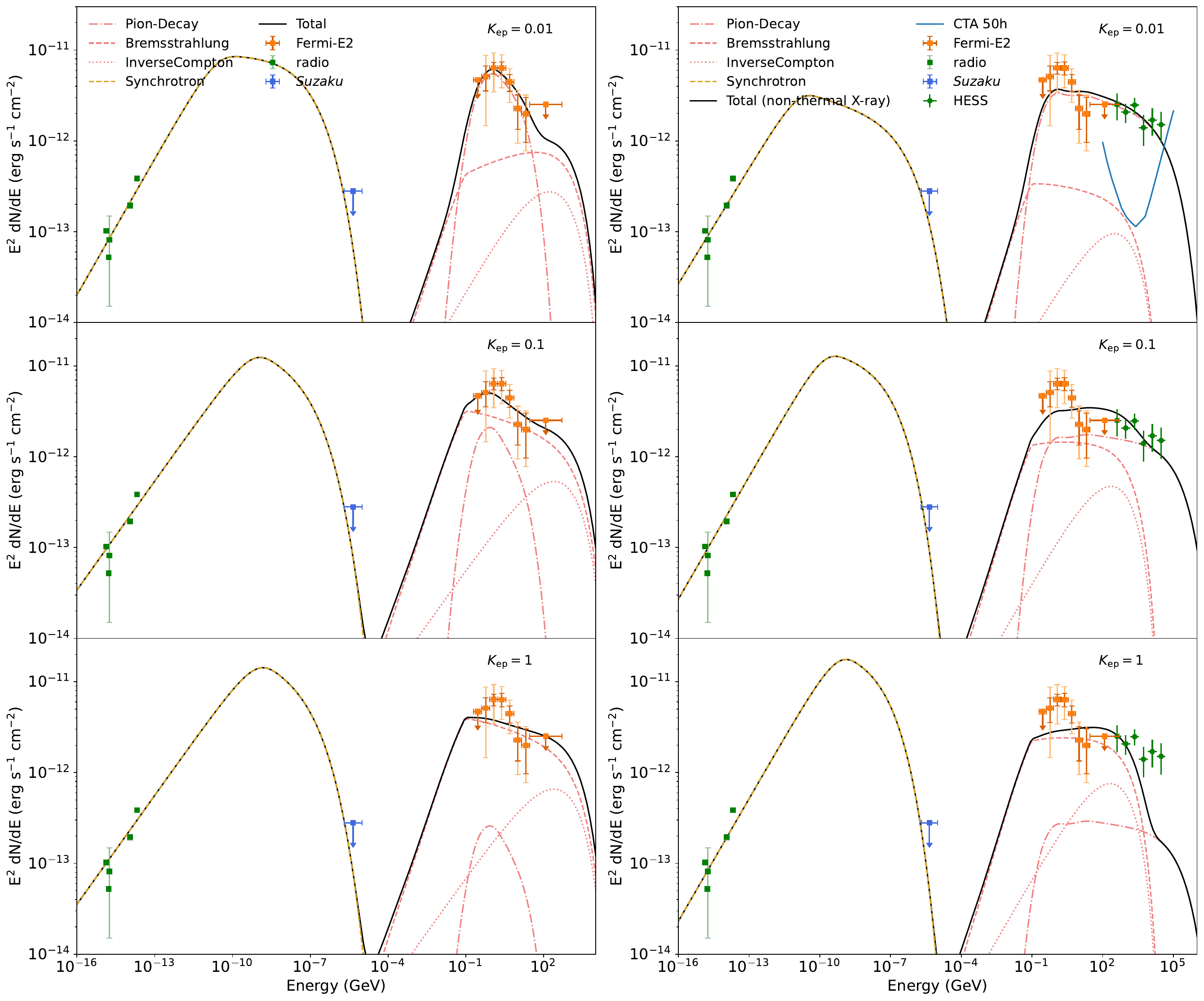}
    \caption{Broadband SEDs of {\sl Fermi}-LAT source E2 with various $K_{\rm ep}$ values. The statistic errors of {\sl Fermi}-LAT data points are in deep colors and the total errors are in light colors. The radio data points of SNR \snr\ are listed in Table~\ref{tab:radio data}, and the X-ray flux is taken from \citet{Kes78_suzaku}.
    %The modeled IC fluxes are too low to appear in the panel \red{with $K_{\rm ep} = 0.01$}.
    {\sl Left:} SED fitting for source E2. 
    {\sl Right:} SEDs for both source E2 and HESS J1852$-$000. 
    The H.E.S.S. data points (green) are adopted from \citet{HESS_catalog}.
    %The solid black lines represent the fits obtained by constraining the electron cutoff energy with the non-thermal X-ray emission, while the dashed black lines correspond to the fits where the X-ray data are treated as upper limits. 
    %The yellow and light-red lines denote the individual components of the non-thermal X-ray emission model.
    The 50 h sensitivity curve of CTA \citep{CTA_sensitivity} is plotted in teal in top right panel.
    %{\sl Left:} \red{SED fitting with $K_{\rm ep} = 0.01$, in which the pion-decay process dominates the GeV emission.}
    %The hadronic model of source E2.
    %{\sl Right:} \red{SED fitting with $K_{\rm ep} = 1$, in which the bremsstrahlung process dominates, suggesting a leptonic origin.}
    %The leptonic model of source E2. The radio data points of SNR \snr\ are listed in Table~\ref{tab:radio data} and the X-ray flux is taken from \citet{Kes78_suzaku}.
    \label{fig: SED}}
    
\end{figure}

We use the \texttt{SED} method of {\small Fermipy} to generate the SED of source E2 in the energy range of 0.2 -- 500\,GeV by applying the maximum likelihood analysis in eight logarithmically spaced energy bins. 
In the fitting process, we free the normalization parameters of the sources within 3$^{\circ}$ from the ROI center and the Galactic background parameters. In the energy bins, when the TS value of source E2 is less than 4, we calculate the 95\% confidence level upper limit of flux. 
The obtained SED is in the left panel of Figure~\ref{fig: SED}. 
For further comparison, we also generate the SED of source PTS1 in the spatial model {\sl 1Ext 2PS} in six logarithmically spaced energy bins, and the results are shown in the right panel of Figure~\ref{fig: PTS1} with the LogP spectral model for its lowest AIC value.
The obtained flux of source PTS1 in 0.2 -- 500 GeV is 1.4$\times\,10^{-11}$ erg cm$^{-2}$ s$^{-1}$, which is 56\% of that of source E2.
Besides the statistic errors, we calculate the systematic uncertainty by fixing the normalization value of Galactic diffuse background at $\pm$6\% of the best-fit value \citep{bkg_pm0.06}.
The maximum deviation between the spectral results obtained with altered background normalizations and those from the nominal fit is taken as the systematic error. We then combine the statistical and systematic uncertainties in quadrature to obtain the final error bars in our spectral analysis.

\section{Discussion}\label{sec:discussion}

%\subsection{Possible Origin of the gamma-ray Emission}

Within the spatial range of the gamma-ray source `E2', there are two high energy sources, SNR Kes78 and PSR J1852-0002g, both of which have the ability to emanate GeV gamma rays. Therefore, we here discuss the potential and role of them in the emission by examining the SED of the `E2' (in {\sl 2Ext} model) or `PTS1' + `PTS2' (in {\sl 1Ext 2PS} model) region

\subsection{Emission from SNR Kes 78}

As revealed above, the extended GeV source E2, with $R_{68}\approx0.31^{\circ}$, projectively covers SNR \snr. 
To analyze the possible origin of the gamma-rays associated with the SNR, we model the gamma-ray spectrum with both hadronic and leptonic mechanisms.
We assume that the accelerated particles have a power-law energy distribution with a high-energy cutoff: 

\begin{equation}
\begin{aligned}
    dN_i/dE =A_i (E_i/ 1\ {\rm GeV})^{-\alpha_i} \exp(-E_i/E_{{\rm cut},i}),
\end{aligned}
\end{equation}
where $i={\rm e, p}$; $E_i$, $\alpha_i$, and $E_{{\rm cut},i}$ are the particle energy, the power-law index, and the cutoff energy, respectively. The normalization $A_i$ is determined by the total energy above 1 GeV, which is converted from the supernova explosion kinetic energy $E_{\rm SN}=10^{51}$ erg with a conversion fraction $\eta$.
We define $K_{\rm ep} = A_{\rm e}/A_{\rm p}$ to control the number ratio of electrons and protons.
For electrons, we also take into account the synchrotron cooling effect that the electron index will be steepened by one for the electron energy above $E_{\rm break} \sim 0.1$ TeV ($B/100\,\mu$G)$^{-2}$ ($t_{\rm age}$/10 kyr)$^{-1}$, where $B$ is the magnetic field strength obtained by SED fittings and $t_{\rm age}$ is the adopted compromised age of SNR \snr\ as $\sim 10$ kyr.
We use PYTHON package Naima \citep{naima} to calculate the SED considering four radiation mechanisms, synchrotron \citep{syn}, non-thermal bremsstrahlung \citep{bremsstrahlung}, inverse Compton \citep[IC,][]{ICS} from the same population of electrons, and pion-decay \citep{pp_decay} processes.
For the IC process, we consider the Galactic disc infra-red background with a temperature of 35~K and an energy density of 0.9\,eV\,cm$^{-3}$ estimated from the interstellar radiation field \citep{seed_photons2} and cosmic microwave background.
The detection of the 1720 MHz OH maser on the eastern boundary of the SNR \citep{OH_maser}, along with studies of the molecular environment \citep{Zhou_kes78_2011}, has revealed that \snr\ is interacting with the surrounding MCs.
Therefore, molecules can be expected to play an important role in the gamma-ray emission.
%For SNR \snr, the detection of an OH maser on its eastern boundary indicates an interaction between the SNR and the surrounding molecular clouds \citep{OH_maser}.
%Previous studies of the molecular environment suggest that Kes 78 is associated with ambient molecular clouds at a $V_{\rm LSR}$ of approximately +81 $\km\ps$ and is evolving within a dense molecular environment \citep{Zhou_kes78_2011}.}
We adopt $\bar{n}({\rm H}) = 2\bar{n}({\rm H_2)} \sim$ 70 cm $^{-3}$ as the average number density of the surrounding target H nuclei for the pion-decay and non-thermal bremsstrahlung processes.
The detailed calculation of molecular parameters in the $R_{68}$ region of source E2 is described in Appendix~\ref{sec: molecular calculation}.

\begin{table}
    \centering
    \caption{SED fitting results.}
    \label{tab: sed}
    \begin{tabular}{ccccccc}
    \hline
    \hline
    Components &  $K_{\rm ep}^a$ & $\alpha$ & $E_{\rm cut, p}$ (TeV) & $E_{\rm cut, e}$ (TeV) & $B$ ($\mu$G) & $\eta$ \\
    \hline
    \multirow{3}{*}{source E2} & 0.01 & 2.0 & 0.1 & 3.6 
                               & 140  & 0.004 \\ 
                               & 0.1  & 2.2 & 0.1$^b$ & 6.6 & 50 & 0.004 \\
                               & 1    & 2.2 & 0.1$^b$ & 6.6 & 45 & 0.002 \\
    \hline
    \multirow{3}{*}{E2 \& HESS J1852$-$000}
                              & 0.01  & 2.15 & 1e3 & 7.4 & 180 & 0.007 \\
                              & 0.1   & 2.1  & 3e3 & 5.2 
                              & 77    & 0.005 \\
                              & 1     & 2.1  & 3e3$^b$ & 4.9 & 50  & 0.002\\
    \hline
    \end{tabular}
    \begin{tablenotes}
        \small
        \item %Table note --- 
        $^a$ Fixed values in SED fittings.\\
        $^b$ Unconstrained because the hadronic process does not dominate.
        %We use $\sim0.1$ TeV as the synchrotron cooling break energy, which is derived by adopting a compromised age of SNR \snr\ as $\sim10^4$ yr and assuming magnetic field strength $B \sim 10^2\mu$G.
        %Actually, the SED fitting of the gamma-rays is not sensitive to this break energy.
        
    \end{tablenotes}
\end{table}

In the SED fitting, we keep the $\alpha_{\rm p} = \alpha_{\rm e}$ and the results mainly depend on $K_{\rm ep}$.
For comparison, we incorporate the TeV data points of HESS J1852$-$000, as reported in \citet{HESS_catalog}, into the SED fitting.
%\red{We fit the SED with both $K_{\rm ep} = 0.01$ and $K_{\rm ep} = 0.1$, which are shown in Figure~\ref{fig: SED}.}
%\red{In the scenario of $K_{\rm ep} = 0.1$, the contribution by the non-thermal bremsstrahlung and the pion-decay processes are comparable while the IC contribution can be ignored in the high-density environment, which gives $\alpha \approx 2.0$, $\eta = 0.001$, $B=38\,\mu$G, and $E_{\rm cut}\approx50$ GeV, as shown in the left panel of Figure~\ref{fig: SED}.
%representing a leptonic-dominated scenario, 
%as shown in the right panel of Figure~\ref{fig: SED}.
%As shown in Figure~\ref{fig: SED}, \red{as $K_{\rm ep}$ increases, leptonic processes gradually become dominant.
As shown in Figure~\ref{fig: SED}, we separately fit the broadband SEDs for the GeV source E2 alone and for the combined spectrum of E2 and TeV source HESS J1852$-$000.
To constrain the electron cutoff energy $E_{\rm cut, e}$, we adopt the high-temperature X-ray component reported in \citet{Kes78_suzaku}, but treated it as purely thermal (because no synchrotron X-rays by the shock font of \snr\ are detected by XMM-Newton \citep{Kes78_XMM}), and its flux is adopted as an upper limit for the non-thermal component. 
%little synchrotron X-rays are detected from SNRs older than 6 kyr).
%If the X-ray component is interpreted as non-thermal, $E_{\rm cut, e}$ can be directly derived from the fit (solid black lines in Figure~\ref{fig: SED}); if, as an extreme alternative, the X-rays are assumed to be purely thermal, they provide only an upper limit on $E_{\rm cut, e}$ (dashed black lines in Figure~\ref{fig: SED}).
%In addition, break energy is obtained by equating the SNR ahe to the electorn cooling timescale.
%synchrotron cooling timescale has been considered to be SNR ageby equating the SNR age to the electron cooling timescale.
%With a compromised age of SNR \snr\ as $\sim10^4$ yr and assumed $B = 100\mu$G, the corresponding cooling break energy is $\sim$125 GeV.
%We adopt the synchrotron cooling energy as $\sim$0.1 TeV \footnote{With the compromised age of SNR \snr\ as $\sim10^4$}.
%and the spectral index above this energy is steepened to $\alpha + 1$. }
The fitting parameters are summarized in Table~\ref{tab: sed}. 
We see that, as $K_{\rm ep}$ increases, the leptonic processes gradually become dominant; however, both hadronic and leptonic scenarios are capable (or essentially capable) of reproducing the observed data points, except for the TeV range in the leptonic dominant scenario (see bottom right panel in Figure~\ref{fig: SED}).
For the Galactic CRs, the value of $K_{\rm ep}$ is typically used as 0.01 \citep[e.g., ][]{DSA0d01, kep=0d01}.
With the adopted $K_{\rm ep}= 0.01$, the GeV gamma-ray emission is dominated by the hadronic process, with bremsstrahlung providing a subdominant contribution.
The magnetic field strength of order $10^2\,\mu{\rm G}$ is typically seen in the SNRs interacting with MCs, such as CTB 37A \citep{CTB37A_highB}, W28 \citep{W28_highB}, and W44 \citep{W44_highB}, as a result of the compression of gas and magnetic field by the shock interaction.
%By adopting one-tenth of the proton cutoff energy $E_{\rm cut, p}$ obtained from the fit to source E2 as the spectral cutoff, applying the relation between cutoff energy and SNR age in \citet{Ecut-to-age} yields an age of $\sim 20$ -- 2000 kyr.?
%which is unusually old for an SNR with detected X-ray emission. 

In the right panels of Figure~\ref{fig: SED}, we simultaneously fit the data of the GeV source E2 and HESS J1852$-$000. 
%With the constrained electron cutoff energy $E_{\rm cut, e}$, the H.E.S.S. data points can be marginally reproduced and even cannot be matched as $K_{\rm ep}$ increases.
As shown in the top and middle right panels (for $K_{\rm ep} = 0.01$ and 0.1, respectively), the H.E.S.S.\ data are  reproduced by the model and dominated by hadronic process.
%whereas the leptonic processes cannot match the TeV data points with the constrained electron cutoff energy $E_{\rm cut, e}$.
However, the TeV source does not coincide with the densest MC at $V_{\rm LSR} = +76$ -- +88 $\km\ps$ associated with \snr; the densest gas is spatially coincident with the GeV source E2 (Figure~\ref{fig: CO map}).
For the $K_{\rm ep} = 1$ case, the dominating leptonic processes cannot match the TeV data points with the constrained electron cutoff energy $E_{\rm cut, e}$.
%The cutoff proton energy $E_{\rm cut, p}\sim1$\,PeV from the fitting of the combined data of source E2 and HESS J1852$-$000 (Table~\ref{tab: sed}) would correspond to an unphysically young age of $\sim 0.2$ -- 30 yr.
Moreover, as shown in Figure~\ref{fig: TSmaps}d, the $R_{68}$ of source E2 only overlaps a little with the extent of the TeV H.E.S.S. source. 
%In addition, the TeV source does not coincide with the densest MC at $V_{\rm LSR} = +76$ -- +88 $\km\ps$ associated with \snr; the densest gas is spatially coincident with the GeV source E2 (Figure~\ref{fig: CO map}).
Therefore, there may be a possibility that the TeV emission observed by H.E.S.S. may originate from an independent source along the line of sight.

\begin{table}
    \centering
    \caption{Details of radio data.}
    \label{tab:radio data}
    \begin{tabular}{ccc}
    \hline
    \hline
        Frequency(GHz) & Flux(Jy) & Reference\\
    \hline
    0.327      & 31.3  & \citet{327MHz} \\
    0.408      & 12.8  & \citet{408and5000MHz} \\
    0.430      & $19.0\pm15.5$  & \citet{430MHz} \\
    2.7        & $7.2\pm0.5$    & \citet{2.7GHz} \\
    5          & 7.7   & \citet{408and5000MHz} \\
    \hline
    \end{tabular}
\end{table}

\subsection{Synchro-curvature Radiation from PSR J1852$-$0002g}\label{sec: SC discuss}

\begin{figure*}
    \centering
    \includegraphics[width=0.65\textwidth]{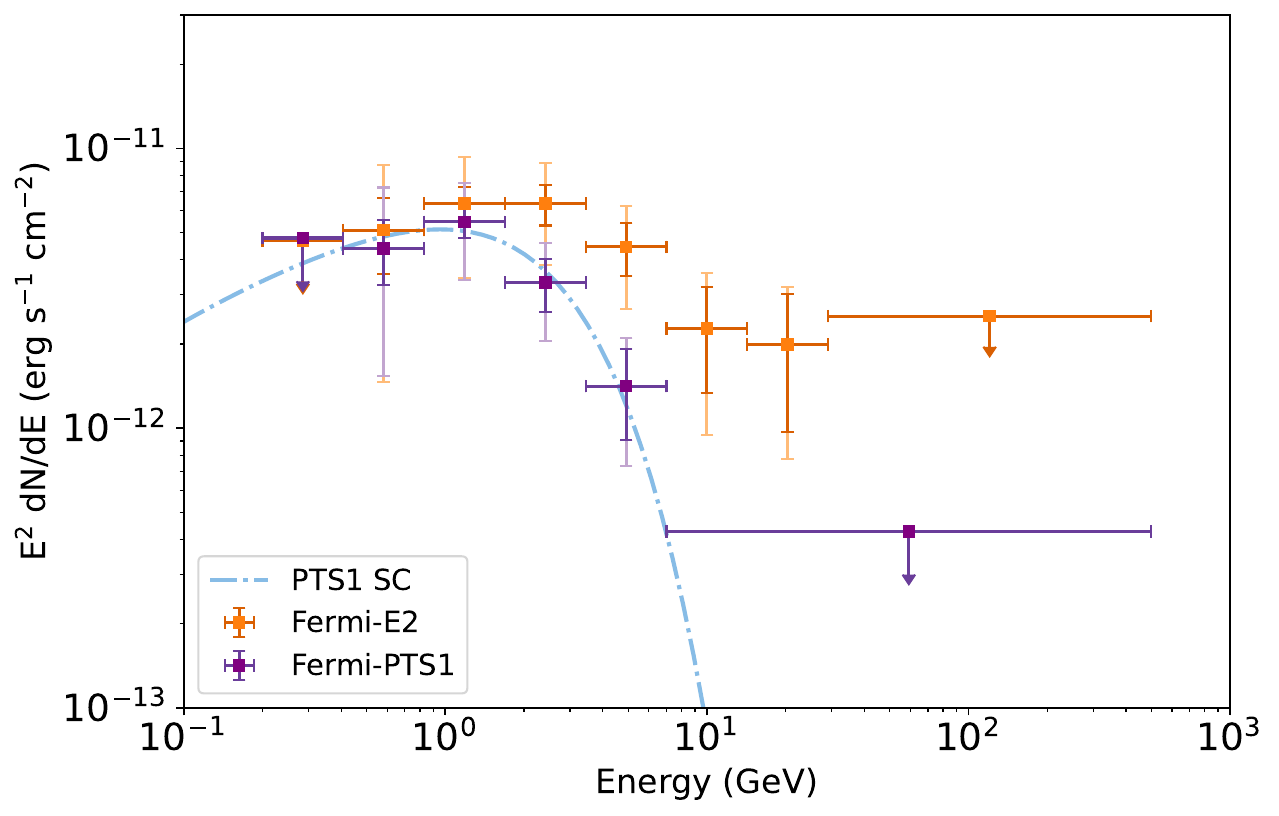}
    \caption{SC model for Fermi-LAT source PTS1 with the data points of source E2 for comparison. The format used to represent the errors of the data points is the same as that in Figure~\ref{fig: SED}.}
    %{\sl Right:} The evolution of Lorentz factor $\Gamma$ and pitch angle $\sin\theta$.}
    \label{fig: PTS1}
\end{figure*}

In Figure~\ref{fig: TSmaps}c, we see that PSR J1852$-$0002g is inside spatial range ($R_{68}$) of the extended source E2 and the 68\% positional uncertainty circle of the test point source PTS1 in the {\sl 1Ext 2PS} model. 
Therefore, a possible gamma-ray contribution from PSR J1852$-$0002g could not be excluded and is worth estimating.
%\red{This PSR, J1852$-$0002g, has firstly been}
We consider that PSR J1852$-$0002g emits synchro-curvature (SC) radiation in gamma-rays, appearing as a point source.
Thus, we have also analyzed the SED of source PTS1, the flux of which is 56\% of source E2 in 0.2 -- 500 GeV (see \S\ref{sec: spec} and Figure~\ref{fig: PTS1}).
Note that PTS2 is excluded from the spatial model when the spectrum of source PTS1 is extracted, ensuring that no emission is attributed to PTS2. 
This provides an estimate of the maximum possible contribution from source PTS1 in this region.
We use the same method as that used in \citet{syn-chr_LHA0341} to treat the SC radiation process and follow the formulae provided by \citet{syn-cur-1996, syn-chr_emission_formulae} to fit the spectrum of source PTS1.
The basic formulae of SC radiation used are described in Appendix~\ref{sec: sc formulae}.
Similar to \citet{syn-chr-fit} and \citet{syn-chr_LHA0341} , we set three free parameters in our calculation: 1) the constant accelerating electronic field $E_{||}$, which is parallel to the magnetic field, determines the energy peak of the spectrum. The range of $E_{||}$ is $\lg(E_{||}$/V m$^{-1}$) = 6.5 -- 9.5. 
2) The length scale $x_0$, which is restricted by the low-energy slope of the spectrum, varies from $x_0/r_{\rm c}$ = 0.001 -- 1 (where $r_{\rm c}$ is the curvature radius).
3) $N_0$, the total number of charged particles in the acceleration region, which is in the range of 10$^{26}$ -- 10$^{32}$ particles.

As in \citet{syn-chr_emission_formulae}, the rest parameters are fixed. 
1) With the period of PSR J1852$-$0002g $P=0.24510$ s \citep{FAST_find_PSR}, we assume that $r_{\rm c}$ is similar to the radius of light cylinder $cP/2\pi\sim1\times10^9$ cm \citep{syn-chr-rc}. 
The upper limit of the integral in Eq.~\ref{equ: dp/de} and the maximum distance of emitting region is $x_{\rm max}$ = $r_{\rm c}$ = 10$^9$ cm. 
2) The magnetic field  strength 
$B$ is assumed as $\sim$ 10$^6$ G, which is set to be constant.
%\citep{syn-chr_emission_formulae}.
3) To numerically solve the momentum equation Eq.~\ref{equ: move}, the initial values of Lorentz factor and pitch angle are set $\gamma=10^3$ and $\theta=45^\circ$, respectively.
%\citep{syn-chr_emission_formulae}. 

%\begin{table}
%    \centering
%    \caption{Fitting results of synchro-curvature radiation.}
%    \label{tab: sc fit result}
%    \begin{tabular}{cccc}
%         \hline
%         \hline
%         Source &log($E_{||}(V m^{-1})$) & x$_0$/r$_{\rm c}$ & N$_0(particles)$ \\
%         \hline
%         E2     &  7.98       & 0.015    &  0.95$\times 10^{31}$ \\
%         PTS1   &  7.35       & 0.060    &  2.30$\times 10^{31}$ \\
%         \hline
%    \end{tabular}
%\end{table}

We use the SC radiation model to fit the spectra of the point source `PTS1' and the fitting results are: $\lg(E_{||}$/V m$^{-1}$) = 7.35, $x_0 / r_{\rm c} = 0.06$ and $N_0 = 2.3\times10^{31}$, with the distance to the PSR adopted as 5.6 kpc \citep{FAST_find_PSR}, as shown in Figure~\ref{fig: PTS1}.
Compared with the results of \citet{syn-chr-fit}, the derived parameters for source PTS1 are generally reasonable.
The length scale $x_0/r_{\rm c}$ is slightly higher than the values reported in their Table 2.
The low-energy slope of the source PTS1 spectrum is relatively steep, indicating a preference for a larger $x_0$.
%This value still falls within a physically reasonable range.
%qOverall, the parameter set provides a satisfactory fit to the observed spectrum of PTS1.
%\red{As shown in the right panel of Figure~\ref{fig: PTS1}, the evolution of $\Gamma$ and $\theta$ illustrates a transition in the electron emission mechanism from synchrotron-dominated to curvature-dominated as the length scale increases.}
%As shown in the right panel of Figure~\ref{fig: PTS1} the evolution of $\Gamma$ and $\theta$ shows the change of electron emission from synchrotron-dominated to curvature-dominated as the length scale increases.}
%when $\Gamma$ approaching asymptotic value $\Gamma_{\rm st} = \left(\frac{3}{2}\frac{E_{||}r^2_{\rm c}}{e}\right)^{1/4} \approx 0.04 r_{\rm c}$.}
%the three stages including the synchrotron-dominated stage, the linear growth of $\Gamma$, and the stabilization stage 
%when $\Gamma$ approaching asymptotic value $\Gamma_{\rm st} = \left(\frac{3}{2}\frac{E_{||}r^2_{\rm c}}{e}\right)^{1/4} \approx 0.04$.
The SED fitting shows that the emission of the potential point source PTS1 could be explained as SC radiation, indicating that the SC radiation produced by PSR J1852$-$0002g could take a part in the detected gamma-rays of source E2.
However, the existence of source PTS1 and its potential contribution in this region remain uncertain and require further observational confirmation.
%Though the SC radiation should be observed as a point source in GeV data, we suggest that there may be the contribution from PSR in this scenario.

\section{Conclusion}
We have analyzed the GeV gamma-ray emission in the region of the SNR \snr\ using $\sim$16.7 years of {\sl Fermi}-LAT observations.
%and investigated its surrounding molecular environment with FUGIN archival \twCO\ and \thCO\ data.
Based on approximately 16.7 years of {\sl Fermi}-LAT data, we find that the catalog sources 4FGL J1852.4+0037e and 4FGL J1851.8$-$0007c are better represented as two extended sources modeled as {\sl 2Ext}.
One of them, designated as E1, is the refitted source of the catalog source 4FGL J1852.4+0037e, with the coordinates R.A. = 283.24$^\circ$, Dec. = 0.89$^\circ$ and $R_{68} =0.45^\circ$.
The other one, designated as E2, is located at R.A.$=282.86^\circ$, Dec.$=-0.11^\circ$ with the $R_{68} = 0.31^\circ$, and is detected with a significance of approximately 15.2$\sigma$ in the 0.2--500 GeV energy range. 
The emission source E2 is well described by a LogP spectral model with spectral index $\Gamma$ = 1.2 and curvature $\beta$ = 0.3. 
The integrated flux in this energy range is $2.5 \times 10^{-11}$ erg cm$^{-2}$ s$^{-1}$, corresponding to a luminosity of $7.6 \times 10^{34}$ erg s$^{-1}$, for a distance of 5 kpc.
%The fitting with ratio $K_{\rm ep}=0.01$ indicates that the GeV emission of source E2 is dominated by hadronic emission.
Given the dense molecular environment surrounding SNR \snr, together with the typical electron-proton ratio $\sim$ 0.01, the hadronic scenario provides a natural explanation for the observed GeV emission.
%Given that the detection of OH maser on the eastern boundary of SNR \snr\ and the SNR is evolving within a dense interstellar environment, Given the molecular environment surrounding SNR \snr, 
%and that no synchrotron X-ray counterpart is detected in archival XMM-Newton data \citep{Kes78_XMM}, 
%a hadronic origin of the GeV emission is favored, with the ratio between electrons and protons $K_{\rm ep}=0.01$.

Although the GeV emission in \snr\ region is best characterized as an extended source E2, it can also be replaced with two point sources. 
One of them, designated as PTS1, is coincident with PSR J1852$-$0002g within the 68\% positional uncertainty circle, indicating a possible gamma-ray contribution from this pulsar. 
The gamma-ray spectrum of source PTS1 is well described by a LogP spectral shape. 
%We apply the SC radiation model to independently fit the SED of source PTS1. 
The SC radiation model provides a satisfactory spectral fit for source PTS1, suggesting that some of the GeV emission from the \snr\ region might possibly originate from the magnetosphere of PSR J1852$-$0002g. 
However, no gamma-ray pulsation has been found from the pulsar in our work, possibly due to limited observational sensitivity or inaccurate ephemeris.

LHAASO has detected emission above TeV in the \snr\ region \citep[i.e., LHAASO J1850$-$0004u, ][]{LHAASO_catalog}.
The emission properties above TeV are expected to be studied and clarified with LHAASO observation and future CTA observation. 
%We, together with LHAASO collaboration, will present the data analysis in the near future.

%\nolinenumbers

\section*{Acknowledgments}
Y.Z.S thanks Wen-Juan Zhong and Jia-Xu Sun for helpful comments.
This work is supported by the National Natural Science Foundation of China (NSFC) under grants 12173018, 12121003, and 12393852. 

\section*{Data Availability}
The Fermi-LAT data underlying this work are publicly available and can be downloaded from \url{https://fermi.gsfc.nasa.gov/ssc/
data/access/lat/}.
We also use the data from FUGIN, FOREST Unbiased Galactic plane Imaging survey with the Nobeyama 45 m telescope, a legacy observation project in the Nobeyama 45 m radio telescope of CO. 
The MeerKAT 1.3 GHz data used in this work are public in \url{https://doi.org/10.48479/3wfd-e270} \citep{MeerKAT}.

\appendix

\section{Molecular parameters}\label{sec: molecular calculation}

\begin{figure}
    \includegraphics[width=1.0\textwidth]{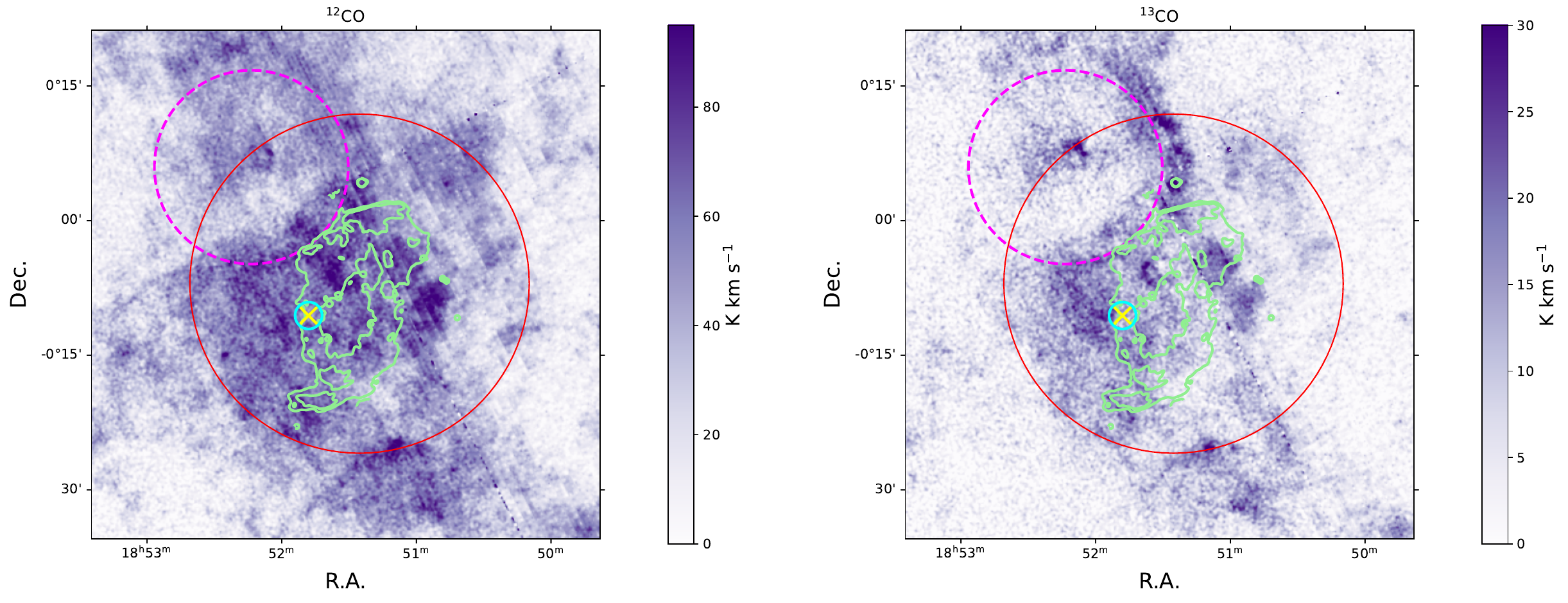}
    \caption{Integrated-intensity greyscale image of \twCO\ (\Jotz) (left panel) and \thCO\ (\Jotz) (right panel) around SNR \snr\ in the $\VLSR$ range +76 -- +88\,km\,s$^{-1}$ (left panel), overlaid with the SARAO MeerKAT Galactic Plane Survey (SMGPS) 1.3\,GHz radio continuum contours of the SNR in green. 
    The red circle marks the extended {\sl Fermi}-LAT GeV source `E2' with the radius $R_{68} \approx 0.31^{\circ}$ around the SNR and the yellow cross is the location of +86.1 $\km\ps$ OH maser on the SNR's eastern boundary, surrounded by the cyan circle with the radius of $0.025^\circ$; the two circled (red and cyan) regions are selected for extracting molecular line profiles that are presented in Figure~\ref{fig: CO lines}.
    The dashed magenta circles are the position of HESS J1852$-$000.}
    \label{fig: CO map}
\end{figure}

\begin{figure}
    \includegraphics[width=1.0\textwidth]{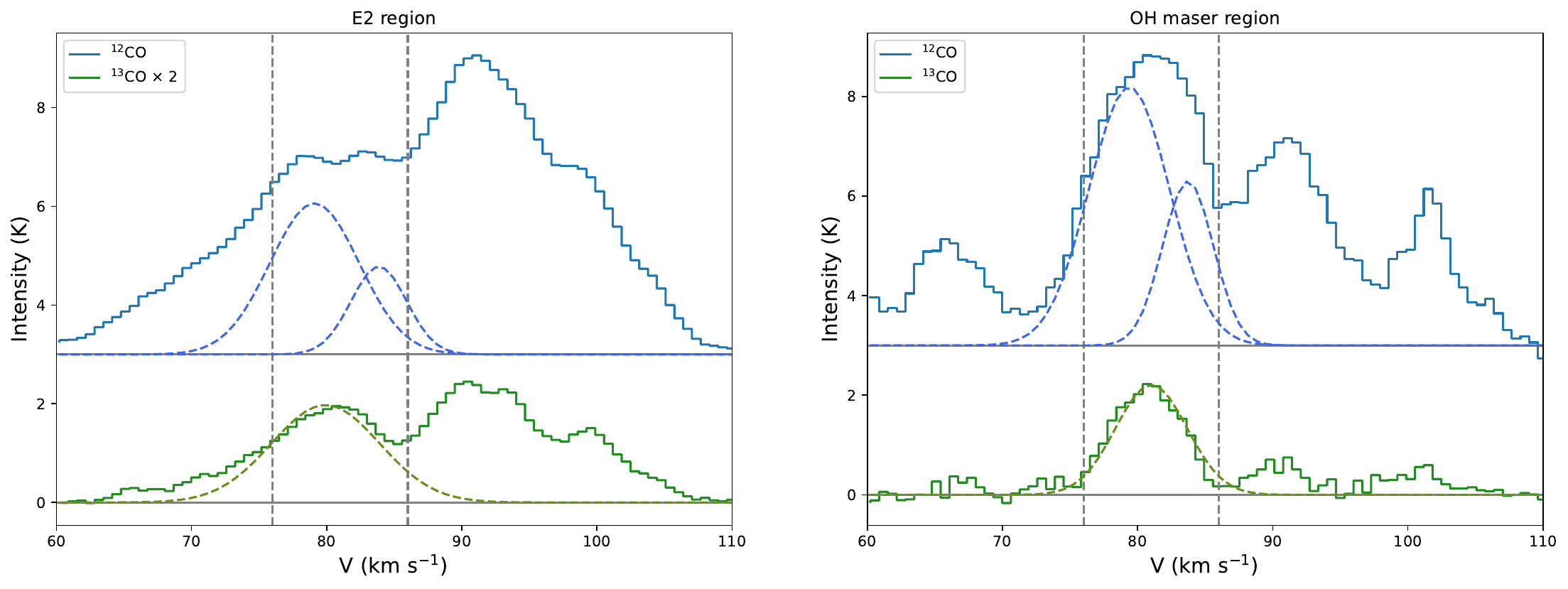}
    \caption{Averaged spectra and multi-Gaussian fitting results in LSR velocity range from +60\,km\,s$^{-1}$ to +110\,km\,s$^{-1}$ for \twCO\ and \thCO\ extracted from `E2' region and OH maser region marked in Figure~\ref{fig: CO map}. The gray vertical lines mark the velocity of +76 $\km\ps$ and +86 $\km\ps$ (the latter is the velocity of the OH maser).}
    \label{fig: CO lines}
\end{figure}

Following the work of molecular environment of SNR Kes78 by \citet{Zhou_kes78_2011}, we use the FUGIN data to calculate the parameters of the surrounding molecular gas that is likely to be involved in the hadronic interaction responsible for the GeV gamma-ray emission.
We extract the averaged CO spectra from the region within $R_{68}\approx 0.31'{\circ}$ for source `E2' and the region around the OH maser \citep{OH_maser}, which are marked in Figure~\ref{fig: CO map}.
According to the Gaussian fitting in the velocity range $\sim$+70 -- $\sim$+90 $\km\ps$, we select the local-standard-of-rest velocity $V_{\rm LSR}$ range +76 -- +86 $\km\ps$ to calculate the molecular column density based on the Gaussian fitting results of \twCO\ and \thCO.
We assume that the \twCO\ lines are optically thick and the \thCO\ lines are optically thin and calculate the column density for the optically thick and thin cases \citep[][also see therein for detailed definition of the symbols used in the two equations below]{Calculate_column_dencity}

\begin{equation}
\begin{aligned}
    N_{\rm thick} =  \frac{3h}{8 \pi^3 \mu^2 J_\mu R_{\rm i}} 
    %\left(\frac{Q_{\rm rot}}{g_{\rm J} g_{\rm K} g_{\rm I}}\right) 
    \left(\frac{k T_{\rm ex}}{hB_{\rm 0}}+\frac{1}{3}\right)
    \frac{\exp{(E_{\rm u}/kT_{\rm ex}})}{\exp{(E_{\rm u}/kT_{\rm ex})} - 1 } \times \int -\ln{\left[1-\frac{T_{\rm R}}{J_{\nu}\left(T_{\rm ex}\right)-J_{\nu}(T_{\rm bg})}\right]^{-1}}dv
\end{aligned}
\end{equation}
and
\begin{equation}
\begin{aligned}
    N_{\rm thin} = \frac{3h}{8 \pi^3 \mu^2 J_\mu R_{\rm i}}\left(\frac{k T_{\rm ex}}{hB_{\rm 0}}+\frac{1}{3}\right) \exp{\frac{E_{\rm u}}{kT_{\rm ex}}}  \times\left(\exp{\frac{h\nu}{kT_{\rm ex}}}-1\right)^{-1}\int\tau_{\rm \nu}dv, 
	\label{equ:HCO+}
\end{aligned}
\end{equation}
The obtained column densities for the E2 region is $N$(\twCO) = $1.2\times10^{17}$ cm$^{-2}$ and $N$(\thCO) = $8.3\times10^{15}$ cm $^{-2}$.

\begin{table}
    \centering
    \caption{Calculated molecular parameters.}
    \label{tab: mc pars}
    \begin{tabular}{cccc}
         \hline
         \hline
         Region & Molecule & $N$(species)$^a$ / cm$^{-2}$ & $\bar{n}({\rm H_2})$ / cm$^{-3}$\\
         \hline
         \multirow{2}{*}{E2 $R_{68}$}
                & \twCO    & $1.2\times10^{17}$       & \multirow{2}{*}{35}   \\
                & \thCO    & $8.3\times10^{15}$                               \\
         \multirow{2}{*}{OH Maser}
                & \twCO    & $2.4\times10^{17}$       & \multirow{2}{*}{700}  \\
                & \thCO    & $1.3\times10^{16}$                               \\
         \hline
    \end{tabular}    
    \begin{tablenotes}
        \small
        \item $^a$ Column density of the specific molecular species.
    \end{tablenotes}
\end{table}

We calculate the $N({\rm H}_2)$ values with the CO-to-H$_2$ conversion factor of velocity integrated brightness temperature 
$N({\rm H_2)} / W(^{12}{\rm CO}) \approx 1.8\times 10^{20} {\rm cm}^{-2}\,\,
                                        {\rm K}^{-1}\,\,{\rm km}^{-1}\,\,{\rm s}$
\citep{12co-to-H2}.
In addition, we also obtain $N({\rm H}_2)$ using abundance ratio $N({\rm H_2 })/N({\rm ^{13}CO}) \approx 6 \times 10^5$ \citep{13CO-to-H2=7e5}.
The results obtained by the above two methods are not much different from each other. 
Assuming that the line-of-sight depth in each region is comparable to the $R_{68}$ for the E2 region and to a radius of 0.025$^\circ$ for the OH maser region, we estimate the 
%mass by $M = 2.8m_{\rm H}N({\rm H}_{\rm 2}){\cal A}$ (where ${\cal A}$ is the projected area for each region) and the 
mean molecular number density $\bar{n}({\rm H_2})$.
The obtained results are listed in Table~\ref{tab: mc pars}.
%In the result, we obtain that the mass of E2 region is $\sim 3.5\times10^5 M_\odot$ and $\nHH \approx$ 35 cm$^{-3}$.

\section{Synchro-curvature radiation model}\label{sec: sc formulae}

The synchro-curvature power radiated by a single particle is given by \citep[see ][]{syn-cur-1996, syn-chr_emission_formulae}:
\begin{equation}
\begin{aligned}
    \frac{{\rm d} P_{\rm SC}}{{\rm d}E} 
    = \frac{\sqrt{3}e^2 \gamma y}{4\pi \hbar r_{\rm eff}}
    [(1+z)\int_y^\infty K_{5/3}(y')dy' - (1-z)K_{2/3}(y)],
    \label{equ:diff power}
\end{aligned}
\end{equation}
where $K_{n}$ are the modified Bessel functions of the second kind of index $n$, $\hbar$ is the reduced Planck's constant, $E$ is the photon energy, $\gamma$ is the Lorentz factor of the electron, $r_{\rm eff}$ is the effective radius,
and $y$ and $z$ are dimensionless quantities related to photon energy and effective radius, respectively, defined in \citet{syn-cur-1996, syn-chr_emission_formulae}.
%\red{More detailed definitions of these quantities could be found in \citet{syn-cur-1996} and \citet{syn-chr_emission_formulae}.}

The evolution of $\gamma$ and $\theta$ (the pitch angle) is obtained by numerically solving the equation for the particle relativistic momentum $\boldsymbol{p}$ (=$\sqrt{\gamma^2-1}mc\hat{p}$):
%We also need to numerically solve the particles' motion equation of relativistic momentum $\boldsymbol{p}$ (=$\sqrt{\Gamma^2-1}mc\hat{p}$, where $\hat{p}$ is the direction vector) to obtain the evolution of $\Gamma$ and $\theta$\red{, which is the pitch angle}:
\begin{equation}
\begin{aligned}
    \frac{d\boldsymbol{p}}{dt} = eE_{||}\hat{b}-\frac{P_{\rm SC}}{v}\hat{p}, 
    \label{equ: move}
\end{aligned}
\end{equation}
where $E_{||}$ is the parallel electric filed, $\hat{b}$ is the direction vector of magnetic field, and $\hat{p}$ is the direction vector of the momentum.

The average SC spectrum throughout the particles' motion trajectory is calculated by the equation
\begin{equation}
\begin{aligned}
    \frac{dP_{\rm tot}}{dE} = \int^{x_{\max}}_0 \frac{dP_{\rm SC}}{dE}\frac{dN}{dx}dx, 
    \label{equ: dp/de}
\end{aligned}
\end{equation}
where $dN/dx$ is the effective weighted distribution of emitting particles as a function of distance $x$ from the injection point toward the observer. It is given by
\begin{equation}
\begin{aligned}
    \frac{dN}{dx} = \frac{N_0 e^{-x/x_{\rm max}}}{x_0(1-e^{-x_{\rm max}/x_0})},
\end{aligned}
\end{equation}
where $N_0=\int^{x_{\max}}_0(dN/dx)dx$ is the normalization and $x_0$ is the length scale of the effective particle distribution.

\bibliography{kes78}{}
\bibliographystyle{aasjournalv7}

\end{CJK*}
\end{document}